\documentclass[aps,prd,reprint,groupedaddress,nofootinbib,onecolumn]{revtex4-2}

\usepackage[utf8]{inputenc} 
\usepackage[T1]{fontenc} 
\usepackage{lmodern}
\usepackage[british]{babel}
\usepackage{graphicx}
\usepackage[pdfusetitle]{hyperref}

\usepackage{xfrac} 
\makeatletter
\g@addto@macro\@floatboxreset\centering
\makeatother

\usepackage[intlimits]{amsmath} 
\usepackage{amssymb} 
\allowdisplaybreaks 

\usepackage[capitalise]{cleveref} 
\creflabelformat{equation}{\textup{#2#1#3}} 
\crefname{section}{Sec.}{Secs.}

\usepackage{booktabs} 
\usepackage{tabularx} 
\usepackage{array} 
\newcolumntype{L}{>{$}l<{$}} 
\newcolumntype{C}{>{$}c<{$}} 

\usepackage[italic]{hepnames} 
\usepackage{siunitx} 
\usepackage{suffix} 

\renewcommand{\Re}{\operatorname{Re}}
\renewcommand{\Im}{\operatorname{Im}}

\newcommand{\Lagrangian}{\ensuremath{\mathcal{L}}\xspace}
\newcommand{\hc}{\text{h.c.}}

\newcommand{\V}[1]{\ensuremath{V_{#1}^{}}} 
\WithSuffix\newcommand\V*[1]{\ensuremath{V_{#1}^*}}

\newcommand{\RVus}{\ensuremath{R(\V{us})}\xspace}
\newcommand{\CHlthree}{C^{(3)}_{H\ell}}
\newcommand{\CHlthreebracket}{\left[C^{(3)}_{H\ell}\right]}
\newcommand{\CHlone}{C^{(1)}_{H\ell}}
\newcommand{\program}[1]{\texttt{#1}}
\newcommand{\flavio}{\program{flavio}\xspace}
\newcommand{\smelli}{\program{smelli}\xspace}
\newcommand{\wilson}{\program{wilson}\xspace}
\newcommand{\DVR}{\Delta^V_R}
\newcommand{\QuantumNumbers}[3]{(\mathbf{#1}, \mathbf{#2}, #3)}

\begin{document}

\author{Matthew Kirk}
\email{matthew.kirk@roma1.infn.it}
\title{Cabibbo anomaly versus electroweak precision tests: An exploration of extensions of the Standard Model}
\affiliation{Dipartimento di Fisica, Università di Roma ``La Sapienza''}
\affiliation{INFN Sezione di Roma, Piazzale Aldo Moro 2, 00185 Roma, Italy}
\date{\today}

\begin{abstract}
There is a newly emerging tension between determinations of \V{us} from different sources (known as the Cabibbo anomaly), clearly demonstrated by the new \RVus observable which is highly sensitive to lepton flavour universality violating effects.
We explore this observable from the perspective of the Standard Model Effective Field Theory and show there is a discrepancy between \RVus and existing electroweak precision observables (EWPO) in a simple single operator dominated scenario.
We explore all possible single particle extensions of the Standard Model that can generate the Cabibbo anomaly effect and show how they cannot simply reconcile the current data.
We further examine the future of EWPO at the ILC or FCC-ee experiments and discuss the effect on the tension of a change in specific EW observables.
\end{abstract}

\maketitle

\tableofcontents

\section{Introduction}

In recent years there has been much excitement and discussion about flavour anomalies relating to possible deviations from the Standard Model (SM) in decays of third generation down-type quarks to leptons \cite{Graverini:2018riw,Bifani:2018zmi}, which were believed to be the the only significant sign of deviations from the SM in data.
However creeping up over the last few years has been evidence of a new discrepancy and it has now become apparent that there is a sizeable deviation from the SM expectation amongst the \V{ud} and \V{us} elements of the Cabibbo-Kobayashi-Maskawa (CKM) matrix.

Recent progress \cite{Seng:2018yzq,Czarnecki:2019mwq,Seng:2020wjq} in calculating radiative corrections to beta decays has shifted the determination of \V{ud}, whose measurement is primarily through such processes.
Combined with the current measurements of \V{us} there is an apparent deviation in the unitarity of the first row of the CKM matrix which is a clear prediction of the SM.
This anomaly is often referred to as the Cabibbo anomaly, since in a two generation model unitarity manifests itself as the quark mixing matrix being determined by a single parameter, the Cabibbo angle.
Since the CKM matrix is unitary by construction in the SM, any deviation must be a sign of beyond the Standard Model (BSM) physics at work -- the only question is what type?

One possible explanation is that there are extra generations of quarks that are yet to be discovered, and that the \(3 \times 3\) quark mixing matrix we know is merely a sub-matrix of a larger, truly unitary one.
The recent deviation, which is the focus of this work, has spurred several works in this direction (see for example \cite{Belfatto:2019swo,Cheung:2020vqm}).
An alternative, which was espoused in \cite{Crivellin:2020lzu} is that the extraction of the CKM elements \V{ud} and \V{us} is affected by new physics (NP) at work in the weak sector, and specifically in the leptonic \PW vertex with a possible lepton flavour universality violating (LFUV) structure.
In that work, the authors constructed a clean observable \RVus, which has increased sensitivity to NP of this form and in particular can distinguish new physics that acts exclusively in the second generation of leptons, i.e.\ in a LFUV way, from a lepton flavour universal (LFU) effect in both electrons and muons.

We extend that work by analysing their minimal modification of the SM in the framework of the Standard Model Effective Theory (SMEFT) (an earlier approach along similar lines can be found in \cite{Cirigliano:2009wk}).
By examining the operators which modify the \PW vertex, we find that the lepton generation dependent effects they hypothesise can only be generated by a single SMEFT operator.
We undertake a global fit to NP in this operator, and find that there exists a tension between the parameter space favoured by the \RVus observable and that favoured by electroweak precision observables (EWPO), which are amongst the most precisely measured constraints on the SM and which dominate the global fit.
The tension is at the level of around \SI{3}{\sigma} and \RVus and EWPO favour opposite sign new physics Wilson coefficients.
We thus investigate more realistic models of BSM physics, where a single new field will introduce several correlated SMEFT coefficients.
Amongst the six possible new particles, we fit each in turn and, excluding a new massive vector state that cannot arise through perturbative unitary extension of the SM, none can provide a significant reduction to the tension in the simple single operator scenario, although we do find that an \(SU(2)_L\) triplet fermion coupled exclusively to muons provides most improvement.

Finally we examine the internals of the EWPO fit and see how a small number of observables are driving the tension between it and \RVus.
Since over the coming years there may be new experiments that carry the potential to massively improve the precision of the electroweak observables, we examine how shifts in the measured values of a very small number of key observables could bring all the data into agreement and provide a consistent signal of new physics.

Our work is laid out as follows: in \cref{sec:cabibbo_anomaly} we explain the background to the Cabibbo anomaly, how it has developed, and how the \RVus observable is well placed to exploit the unitarity of the CKM matrix.
Following that in \cref{sec:SMEFT} we provide our SMEFT analysis of the situation and what the current Cabibbo anomaly corresponds to in terms of a SMEFT Wilson coefficient.
Next we describe and show the results of our global fit in \cref{sec:global_fit} and explore the tension that appears.
In \cref{sec:BSM} we make our complete exploration of the space of BSM models for the proposed change to the \PW vertex, and that section ends with the specific fits for each scenario.
Our last piece of analysis concerns the future of EWPO and is given in \cref{sec:future}, while we summarise our findings in \cref{sec:summary}.

\section{Cabibbo anomaly in CKM unitarity}
\label{sec:cabibbo_anomaly}

An important prediction of the SM is that the CKM matrix is unitary, being constructed from the product of the two unitary matrices which act to diagonalise the up and down quark Yukawas.
This means that if we measure all the individual CKM elements separately, the CKM matrix is overdetermined and we can use the unitarity condition as a consistency check.
\footnote{In Ref.~\cite{Grossman:2019bzp} the authors argue for testing the equality of the Cabibbo angle rather than first row unitarity as being more statistically robust once multiple measurements of the two CKM elements \V{ud} and \V{us} are considered.}
One can write down many different unitarity conditions, which split into two types:
\begin{enumerate}
	\item The famous ``unitarity triangles'', which are graphical representations of three complex numbers summing to zero (e.g.\ \(\V{ud} \V*{ub} + \V{cd} \V*{cb} + \V{td} \V*{tb} = 0\)).
	\item Sums over absolute squares of row or columns being equal to 1.
\end{enumerate}
The Cabibbo anomaly is an apparent violation of a condition of the second type, namely the sum of absolute squares of the first row of the CKM matrix equalling unity:
\begin{equation}
\label{eq:first_row_unitarity}
|\V{ud}|^2 + |\V{us}|^2 + |\V{ub}|^2 \overset{?}{=} 1
\end{equation}
where the question mark indicates this is a condition which can be tested, and as we describe now, seems to be violated.

To examine the violation of the relation \cref{eq:first_row_unitarity}, we define a new quantity
\begin{equation}
\label{eq:DeltaCKM_defn}
\Delta_\text{CKM} \equiv 1 - |\V{ud}|^2 - |\V{us}|^2 - |\V{ub}|^2
\end{equation}
which is zero in the SM.
(For the purposes of this work, we use the value \(|\V{ub}| = \num{4e-3}\) \cite{Zyla:2020zbs,PDG:CKM_review} and neglect its uncertainty as it is so small as to make no material difference.)

\V{ud} can be measured or extracted from several different experimental processes -- super-allowed atomic beta decays (where super-allowed refers to there being no change to the angular momentum or parity of the nucleus), measurements of the neutron lifetime, or of charged pion decay.
Of these, super-allowed atomic beta decays are currently around an order of magnitude more precise than the others, and so are the best way to determine \V{ud}.
There have recently been significant changes in this method, which require a more detailed discussion which is done in \cref{sub:CKM_current}. 

\V{us} can be found from semi-leptonic kaon decays \HepProcess{\PK \to \Ppi \ell \nu} (often referred to as \(K_{\ell 3}\)) where \(\ell\) is either an electron or muon, or from a ratio of the purely muonic decays \HepProcess{\PK \to \mu \nu} and \HepProcess{\Ppi \to \mu \nu} (similarly known as \(K_{\mu 2}\)).
In the \(K_{\ell 3}\) case, the experimental measurement of the semi-leptonic branching ratios determines the product \(|\V{us}| f_+(0)\), where \(f_+(0)\) is one of the \(\PK \to \Ppi\) form factors at zero momentum transfer, and which is independently determined from lattice QCD calculations.
Looking instead at both kaon and pion decays, an experimental measurement gives the ratio \(|\V{us} / \V{ud} | \times f_K / f_\pi\), and again by independently calculating the ratio of decay constants using lattice QCD, the CKM elements are extracted.
(For further details on the \V{us} determinations and lattice QCD calculations see the review \cite{Moulson:2017ive}.)

In this work, we take the Particle Data Group (PDG) average for the \(K_{\mu 2}\) extraction, \(\V{us} = \num{0.2252 \pm 0.0005}\) \cite{Zyla:2020zbs,PDG:CKM_review}, since this is larger than the \(K_{\ell 3}\) result and hence gives a conservative result for the unitarity deviation.
\footnote{Using the PDG \(K_{\ell 3}\) result, all the significances in \cref{tab:CKM_unitarity_values} would increase by around 1.5 standard deviations.}

\subsection{Current status}
\label{sub:CKM_current}

An extraction of \V{ud} from super-allowed beta decays needs knowledge of various nuclear parameters, of which the quantity \(\DVR\) (which contains the nucleus independent electroweak radiative corrections) is the most important as it is the least precisely known.
A schematic formula for these decays looks like
\begin{equation}
\label{eq:beta_decay_schematic}
\mathcal{F}t = \frac{K}{2 G_F^2 V_{ud}^2 \times (1+\DVR)} \,,
\end{equation}
where \(\mathcal{F}t\) is the measured quantity multiplied by certain well-known corrections relating to the particular nuclear decay and \(K\) is simply the dimensionful constant \(2 \pi^3 \ln 2 / m_e^5\).
By taking the final ingredient \(G_F\) from a measurement of the muon lifetime one can find \V{ud} from the beta decay measurement.
A more detailed description of the theoretical predictions for super-allowed beta decays can be found in \cite{Hardy:2014qxa,Hardy:2018zsb}, but we leave it at this schematic level as it is not relevant for the main results of this paper.
For a long time, the state-of-the-art result for \(\DVR\) was from 2006 \cite{Marciano:2005ec}, where the authors calculated \(\DVR = \num{0.02361(38)}\) which implies \(\V{ud} = \num{0.97420(21)}\) (using 2018 experimental beta decay data \cite{Hardy:2018zsb}).
As can be seen in \cref{tab:CKM_unitarity_values}, this value of \V{ud} means the CKM unitarity violation parameter \(\Delta_\text{CKM}\) is consistent with zero, and hence there was no sign of any problems with the SM.

In the last few years, there has been much work done on this quantity using different calculational tools, which have improved the precision \emph{and} shifted the central value upwards, leading to a deficit in \V{ud}.
First in \cite{Seng:2018yzq} the authors used dispersion relation methods and produced the result \(\DVR = \num{0.02467(22)}\) and \(\V{ud} = \num{0.97370(14)}\).
This method has been widely used for other calculations (see Refs.~7 through 20 in that work).
Another calculation using an alternative method has been done in \cite{Czarnecki:2019mwq}, wherein the authors found \(\DVR = \num{0.02426(32)}\) and \(\V{ud} = \num{0.97389(18)}\).
Finally, a further analysis this year \cite{Seng:2020wjq} gave yet another indication of larger radiative corrections: \(\DVR = 0.02477(24)\) giving \(\V{ud} = \num{0.97365(15)}\).
In light of all these new results, we calculate a weighted average of the new calculations:
\begin{equation}
\label{eq:weighted_average}
\DVR = \num{0.02462(14)} \; \Rightarrow \; \V{ud} = \num{0.97373(9)} \,,
\end{equation}
which provides an improvement over the precision of the older 2006 result of almost a factor of three.
Using this value of \V{ud}, and the PDG average for \V{us} and \V{ub} mentioned at the beginning of this section, we find a deviation from first row unitarity of
\begin{equation}
\Delta_\text{CKM} = \num{1.12 \pm 0.28 e-3} \,.
\end{equation}
This shows a deviation from the SM null result of \SI{3.9}{\sigma}, an anomaly which is the main motivation for this paper.
\footnote{Such a large anomaly has been confirmed by various other analyses \cite{Grossman:2019bzp,Belfatto:2019swo,Tan:2019yqp}.}

\begin{table*}
\begin{tabular}{@{}S[table-format=1.3(3)]S[table-format=1.5(5)]rS[table-format=1.2(2)]r@{}}
\toprule
\(\DVR \times 10^2\) & \(\V{ud}\) & Source & \(\Delta_\text{CKM} \times 10^3\) & Significance \\
\midrule 
2.361(38) & 0.97420(21) & MS \cite{Marciano:2005ec,Hardy:2018zsb} & 0.16 \pm 0.52 & \SI{0.3}{\sigma} \\
2.467(22) & 0.97370(14) & SGPR \cite{Seng:2018yzq} & 1.18 \pm 0.35 & \SI{3.3}{\sigma} \\
2.426(32) & 0.97389(18) & CMS \cite{Czarnecki:2019mwq} & 0.81 \pm 0.42 & \SI{1.9}{\sigma} \\
2.477(24) & 0.97365(15) & SFGJ \cite{Seng:2020wjq} & 1.27 \pm 0.37 & \SI{3.5}{\sigma} \\
\addlinespace
2.462(14) & 0.97373(9) & & 1.12 \pm 0.28 & \SI{3.9}{\sigma} \\
\bottomrule
\end{tabular}
\caption{\(\DVR\) and \V{ud} values from various sources and the deviation from top row unitarity they imply.
The final row uses a weighted average of the three recent determinations.
We use the current PDG average for \V{us} stated above everywhere.}
\label{tab:CKM_unitarity_values}
\end{table*}

\subsection{\RVus observable}

The \RVus observable, as introduced in \cite{Crivellin:2020lzu}, nicely exploits the correlation between \V{ud} and \V{us} as implied by the unitarity of the CKM matrix (remember we are proceeding under the assumption that the non-zero value of \(\Delta_\text{CKM}\) is a sign of LFUV rather than a sign that the \(3 \times 3\) CKM matrix is non-unitary due to being a sub-matrix of a larger matrix arising from extra generations of quarks), combined with the large hierarchy between them, to give a better sensitivity to new physics effects.
Put simply, the unitarity condition means that a larger (or smaller) \V{ud} value must imply a smaller (or larger) \V{us} value, with the two changes equal and opposite.
Then since \V{ud} is much greater than \V{us}, the change to \V{us} is relatively much bigger and so we gain a large sensitivity to NP in \V{us}.
\footnote{One can think of this result in terms of \V{ud} and \V{us} both being determinations of a single parameter, the Cabibbo angle \(\theta_c\), and the difference in size between \(\sin \theta_c\) and \(\cos \theta_c\) when \(\theta_c\) is small.}

In \cite{Crivellin:2020lzu} they defined the observable \RVus as
\begin{equation}
\label{eq:RVus_defn}
\RVus \equiv \frac{V_{us}^{K\mu2}}{V_{us}^\beta} \equiv \frac{V_{us}^{K\mu2}}{\sqrt{1 - |V_{ud}^\beta|^2 - |\V{ub}|^2}} \,,
\end{equation}
where the superscripts indicate the process through which the CKM elements are measured.
By then making a naive change to the \PW leptonic vertex
\begin{equation}
\label{eq:naive_W_change}
\Lagrangian \supset \frac{g}{\sqrt{2}} W^-_\mu \bar{\ell}_i \gamma^\mu P_L \nu_j \delta_{ij} \to \frac{g}{\sqrt{2}} W^-_\mu \bar{\ell}_i \gamma^\mu P_L \nu_j ( \delta_{ij} + \varepsilon_{ij} ) \,,
\end{equation}
where \(i,j\) refer to lepton generations, and \(\varepsilon\) is diagonal (\(\varepsilon_{ij} = 0\) if \(i \neq j\)) and small (\(\varepsilon^2 = 0\)), 
two things change.
First, the theoretical expression for the muon lifetime
\begin{equation}
\label{eq:muon_lifetime}
\frac{1}{\tau_\mu} = \frac{G_F^2 m_\mu^5}{192\pi^3} \to \frac{G_F^2 (1 + \varepsilon_{ee} + \varepsilon_{\mu\mu})^2 m_\mu^5}{192\pi^3} \,,
\end{equation}
and secondly the beta decay master formula (\cref{eq:beta_decay_schematic}) where
\begin{equation}
G_F \V{ud} \to G_F \V{ud} (1 + \varepsilon_{ee}) \,.
\end{equation}
Combining these two, we see that the numerical value being extracted from super-allowed beta decays is in fact the combination
\begin{equation}
V_{ud}^\beta \equiv \sqrt{\frac{K}{2 \mathcal{F}t (1+\DVR)}} \sqrt{\frac{\tau_\mu m_\mu^5}{192 \pi^3}} = \V{ud} (1 - \varepsilon_{\mu\mu}) \,.
\end{equation}
Plugging this back into \cref{eq:RVus_defn} and expanding in the small parameter \(\varepsilon\) gives the nice result:
\begin{equation}
\RVus \approx 1 - \left(\frac{\V{ud}}{\V{us}}\right)^2 \varepsilon_{\mu\mu} \approx 1 - 20 \varepsilon_{\mu\mu} \,,
\end{equation}
which clearly shows the large enhancement in sensitivity caused by the hierarchy between \V{us} and \V{ud}.
Note that within this relation it does not matter which values we pick for \V{ud} and \V{us} out of all the values given above, since the absolute differences between the various determinations are small compared to the relative size of the ratio \(\V{ud} / \V{us}\).
\footnote{To be precise, the enhancement factor is \num{18.7 \pm 0.1} which is what is used to produce the numerical result in \cref{eq:my_RVus_number}.
This error takes into account the actual uncertainties on the determinations of \V{ud} and \V{us}, and the range of different \V{ud} values given in \cref{tab:CKM_unitarity_values}.}
Using our weighted average for \V{ud} from beta decays (\cref{eq:weighted_average}) we calculate
\begin{equation}
\label{eq:my_RVus_number}
\RVus = \num{0.9891 \pm 0.0027} \; \Rightarrow \; \varepsilon_{\mu\mu} = \num{0.58 \pm 0.15 e-3} \,,
\end{equation}
again showing a large deviation of just under \SI{4}{\sigma} from the SM expectation \(\RVus = 1\).

\section{The SMEFT perspective}
\label{sec:SMEFT}

In order to analyse the Cabibbo anomaly as set out in the previous section from a more rigorous point of view, we work within the framework of the SMEFT.
The approach of this EFT is to augment the SM with all possible higher dimensional operators that are invariant under the \(SU(3)_c \otimes SU(2)_L \otimes U(1)_Y\) gauge symmetry of the Standard Model, and is valid on the basis that all the extra degrees of freedom beyond the Standard Model are much heavier than the electroweak scale such that an expansion in \(v^2 / M_\text{NP}^2\) is valid.
We work entirely within the dimension-6 SMEFT, so that only non-renormalisable operators of dimension 6 plus the dimension-5 Weinberg operator \cite{Weinberg:1979sa} are added -- our normalisation convention is that the SMEFT Lagrangian takes the form
\begin{equation}
\Lagrangian_\text{SMEFT} = \Lagrangian_\text{SM} + \sum_i C_i (\mu) Q_i (\mu) 
\end{equation}
such that our Wilson coefficients \(C_i\) are dimensionful with mass dimension -2 (or -1 for the Weinberg operator).
We work in units where they have dimensions of \si{\GeV^{-2}} (or \si{\GeV^{-1}}).
The first complete and non-redundant basis for the dimension-6 SMEFT, now known as the ``Warsaw basis'', was given in \cite{Grzadkowski:2010es} and is the basis we use in our work.

There is a technicality associated with SMEFT analysis which is the choice of input parameter scheme.
Depending on what set of measurable inputs you take as relating directly to certain theory parameters, the output expressions can differ (this is due to the overcompleteness of the basis of potential inputs in the SM electroweak sector).
Traditionally the choice was to use \(\{\alpha_\text{EM}, M_Z, G_F, \ldots\}\) as the numerical inputs, but recently there has been a shift towards using \(M_W\) instead of \(\alpha_\text{EM}\) \cite{Brivio:2017bnu}.
However, the software we use for our numerical fits uses the \(\{\alpha_\text{EM}, M_Z, G_F\}\) scheme and so we will work only within this scheme as well.

With these background details specified and out of the way, we move on to considering the \RVus observable within the context of the SMEFT.

\subsection{SMEFT for \RVus}

In order to examine all the possible ways in which SMEFT operators can change the leptonic \PW vertex, we start by considering the full set of operators in the Warsaw basis.
The form of the modification seen in \cref{eq:naive_W_change}, in particular the appearance of lepton generation indices which allow for LFUV NP, narrows down the set considerably by restricting us to consider only SMEFT operators that also have lepton generation indices.
Combined with the requirement to modify the interaction with left handed leptons, we are then left with a single operator -- the \(SU(2)_L\) triplet operator
\begin{equation}
\left[Q_{H\ell}^{(3)}\right]_{ij} = \left( \PH^\dagger i \overset{\leftrightarrow}{D^a_\mu} \PH \right) \left( L_i \sigma^a \gamma_\mu L_j \right) \,.
\end{equation}
(See \cref{app:smeft_definitions} for a full explanation of our notation, which as mentioned above also matches that in \cite{Grzadkowski:2010es}.)
This generates the exact effect of \cref{eq:naive_W_change} after the Higgs gains a vacuum expectation value, with LFUV effects possible if the different components \(\CHlthreebracket_{ij}\) are not identical.

For the moment, we proceed assuming there is only new physics in this single operator (while keeping in mind that this is obviously a basis dependent statement -- later in \cref{sec:BSM} we study more realistic scenarios, inspired by specific BSM possibilities).
The EW gauge boson interactions with leptons are changed in the following way:
\begin{widetext}
\begin{align}
\Lagrangian_\text{SM} \to \Lagrangian \supset &\phantom{+} \frac{g}{\sqrt{2}} W^-_\mu \bar{\ell}_i \gamma^\mu P_L \left( \delta_{ij} + v^2 \CHlthreebracket_{ij} \right) \nu_j + \hc \\
&+ \frac{g}{2 c_\theta} Z_\mu \bar{\ell}_i \Bigg[ \phantom{+} \gamma^\mu P_L \left( \delta_{ij} (-1 + 2 s_\theta^2) - v^2 \CHlthreebracket_{ij} \right) + \gamma^\mu P_R \left( \delta_{ij} (2 s_\theta^2) \right) \Bigg] \ell_j \\
&+ \frac{g}{2 c_\theta} Z_\mu \bar{\nu}_i \gamma^\mu P_L \left( \delta_{ij} + v^2 \CHlthreebracket_{ij} \right) \nu \,.
\end{align}
\end{widetext}
Here we see how the modifying the \PW leptonic vertex in a gauge invariant way gives rise to unavoidable and correlated effects in the \PZ leptonic vertices as well.

We also note the simple direct relationship between the SMEFT coefficient \(\CHlthree\) and the \(\varepsilon\) parameter introduced in \cite{Crivellin:2020lzu} and \cref{eq:naive_W_change}: \(\varepsilon_{ij} = v^2 \CHlthreebracket_{ij}\), which means we can write the \RVus observable as
\begin{equation}
\RVus = 1 - \left(\frac{\V{ud}}{\V{us}}\right)^2 v^2 \CHlthreebracket_{22} (\mu \sim \SI{1}{\GeV})
\end{equation}
In the above equation, we have explicitly written the scale dependence, since \RVus is determined from measurements around the scale \(\mu \sim \SI{1}{\GeV}\).
For the rest of this paper, we will work with and quote SMEFT coefficient numerics at the scale \(\mu = \SI{1}{\TeV}\) unless otherwise specified.
At the level of accuracy we are considering, the renormalisation group effects are purely multiplicative and so we include these automatically using results from \wilson \cite{Aebischer:2018bkb}.
In this fashion, we therefore state here that our \RVus result in \cref{eq:my_RVus_number} corresponds to a result for the SMEFT coefficient of
\begin{equation}
\label{eq:my_CHl322_number}
\CHlthreebracket_{22} (\mu = \SI{1}{\TeV}) = \SI{1.17 \pm 0.30 e-8}{\GeV^{-2}} \,.
\end{equation}

\section{Global fit}
\label{sec:global_fit}

In this section we perform a global fit to data using the software package \smelli \cite{Aebischer:2018iyb} v2.0.0 \cite{smelli_2.0.0}, which contains 399 different observables as of that version.
\footnote{\smelli is based on \flavio \cite{Straub:2018kue} and \wilson \cite{Aebischer:2018bkb}.}
For the purposes of our fit, we take a two-dimensional parameter space of \(\CHlthreebracket_{11}\) and \(\CHlthreebracket_{22}\).
The 22 element is what enters the observable \RVus, while in addition both elements enter into many observables through the change to the muon lifetime from which \(G_F\) is measured (see \cref{eq:muon_lifetime}) and so both are well placed to be constrained by a global fit.
Fitting to both elements allows us to distinguish between data favouring a purely muonic effect, a LFU effect with \(\CHlthreebracket_{11} = \CHlthreebracket_{22}\) or something else in between.
(The third element \(\CHlthreebracket_{33}\) will be mostly constrained by \Ptau physics which are poorly measured, and so we assume zero effect here for simplicity.
We also neglect off-diagonal elements as these are very strongly constrained by the experimental results for lepton flavour violating observables such as \HepProcess{\mu \to e e e} \cite{Bellgardt:1987du} or \HepProcess{\PZ \to e \mu} \cite{Aad:2014bca} except in a very specific case, which we discuss briefly in \cref{sec:BSM}.)

\begin{figure}
\includegraphics[width=0.65\textwidth]{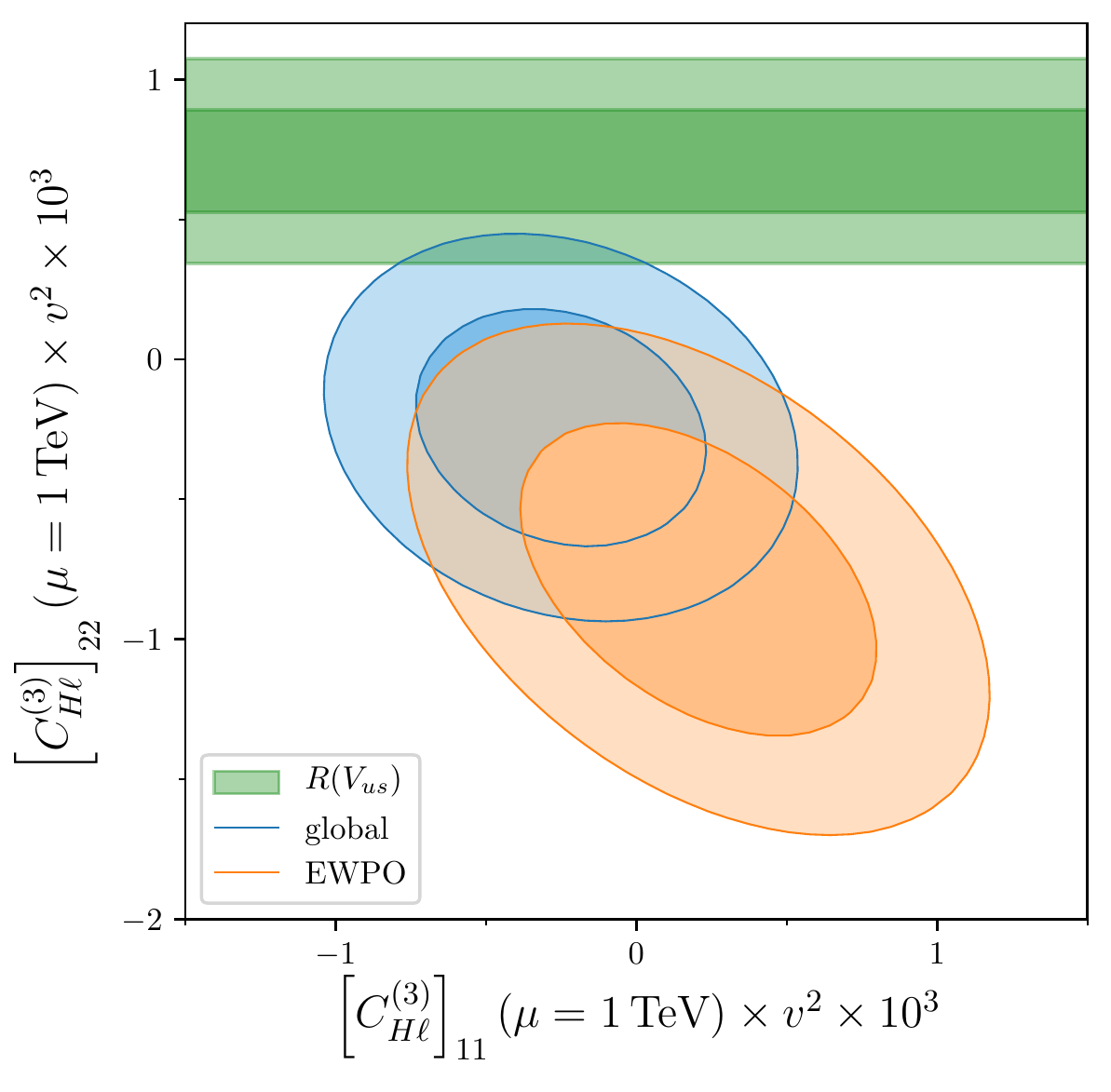}
\caption{Fits to the entire set of observables included in \smelli (excluding only those relating to beta decay) (blue) and to just the subset of EWPO (orange).
The green region corresponds to our result for the \RVus observable \cref{eq:my_CHl322_number}. For each, the dark and light regions correspond to the \SIlist{1;2}{\sigma} allowed regions respectively.}
\label{fig:global_and_EWPT_vs_RVus}
\end{figure}

Before discussing the fit, we note one point: for the global fit, we actually exclude observables relating to beta decay, as in this way a clearer comparison can be made.
The results of our fit are shown in \cref{fig:global_and_EWPT_vs_RVus}, where we show three distinct regions.
In green is best fit for the SMEFT coefficient value as found directly by the \RVus observable and given in \cref{eq:my_CHl322_number}, in blue is the best fit region for the global fit, and in orange is the best fit region for a subset of the observables corresponding to electroweak precision observables.%
\footnote{This subset of observables is listed in \cref{tab:smelli_obs_EWPT}, along with the experimental measurements and theory implementations used in \flavio.}
In all cases (and going forward in all subsequent figures), the darkest region corresponds to a \SI{1}{\sigma} allowed range, and the lighter shaded region is \SI{2}{\sigma} allowed.
We have singled out the subset of EWPO from the global fit since they are the strongest constraints on these coefficients -- the shift upwards between the EWPO and global fit is driven by a very broad preference for LFU from charged current decays.

The global fit shows that (excluding beta decay observables) the current experimental evidence points towards a small negative effective in both \(\CHlthreebracket_{11}\) and \(\CHlthreebracket_{22}\), albeit in a non-significant way.
Numerically, we find 
\begin{equation}
\begin{aligned}
\CHlthreebracket_{11} &= \SI{-0.41 \pm 0.53 e-8}{\GeV^{-2}} \,, \\
\CHlthreebracket_{22} &= \SI{-0.40 \pm 0.46 e-8}{\GeV^{-2}} \,,
\end{aligned}
\end{equation}
with a weak negative correlation coefficient of \(-0.2\).
(We note that these results agree with other recent global SMEFT fits to single operator scenarios in the literature \cite{deBlas:2017wmn,Ellis:2018gqa,Dawson:2020oco}.)
We see here already there are almost \(3\) standard deviations between the central values of beta decay result from \RVus and all other observables in the global fit.
Looking now at the EWPO fit we see an even clearer discrepancy straight away from \cref{fig:global_and_EWPT_vs_RVus}.
Electroweak measurements alone favour a \emph{negative} contribution to \(\CHlthreebracket_{22}\) while providing no argument for a non-zero contribution in the 11 element, in contrast to the \emph{positive} contribution needed to explain the apparent CKM unitarity violation.
Numerically, the region shown corresponds to
\begin{equation}
\label{eq:EWPO_bf}
\begin{aligned}
\CHlthreebracket_{11} &= \SI{-0.35 \pm 0.64 e-8}{\GeV^{-2}} \,, \\
\CHlthreebracket_{22} &= \SI{-1.3 \pm 0.61 e-8}{\GeV^{-2}} \,,
\end{aligned}
\end{equation}
with a stronger correlation of \(-0.45\).
The difference between EWPO alone and \RVus is now much larger, standing at more than three and a half standard deviations.
We believe that such a large internal tension between different sets of observables is worth investigating, particularly to see whether it holds up in more realistic scenarios where several SMEFT coefficients are generated.
One possibility would be to enlarge the dimension of our fit to encompass a larger set of SMEFT operators which are relevant to EWPO, allowing all the Wilson coefficients to vary freely.
Instead however we choose to study simplified UV models, which generate several SMEFT coefficients which are related in a model specific way, as this gives insight into realistic models of BSM physics with a minimum of free parameters.
In particular, as we will soon see, the sign of the coefficient is very important in disentangling possible fermionic BSM explanations, further motivating our next section.

\section{BSM models}
\label{sec:BSM}

As demonstrated from the fit and numerics in the previous section, there is a tension between the regions that are preferred by the current set of EWPO and that of the clean observable \RVus, when looking at a minimal possibility of new physics in the SMEFT coefficient \(\CHlthree\) alone.
Since in a realistic extension of the SM with new fields it is unlikely that this single operator is generated and nothing else, we choose to examine all the possible BSM particles in the following way.
\begin{enumerate}
	\item First, pick out all new fields that generate \(\CHlthree\).
			We do this using the results of \cite{deBlas:2017xtg}, in which the authors provide a simple dictionary between all possible extensions of the SM
			\footnote{In fact, they impose some restrictions on the space of new particles they consider in order to ensure the SMEFT is a good description. Perhaps more importantly, they consider on non-anomalous extensions of the SM, which means only allowing for fermions which are vector-like (i.e.\ non-chiral) under the SM gauge group.}
			and the effective Wilson coefficients they generate after being integrated out at tree-level.
	\item Having identified the relevant subset, we then see what other SMEFT coefficients are generated, using only the couplings necessary for \(\CHlthree\) to be non-zero. This is necessary since some of the new particles have several independent coupling constants, and so generate coefficients which are not able to be related to the triplet operator we are interested in.
	\item Finally (in \cref{sub:BSM_fit}) we fit to EWPO with the specific correlated coefficients corresponding to each new physics scenario and see whether the tension seen in the simple case is relaxed.
\end{enumerate}

We identify six single particle extensions of the SM that generate the operator \(Q_{H\ell}^{(3)}\) after being integrated out, of which four are fermions and two are vector bosons.
These are summarised in \cref{tab:BSM_models} along with their quantum numbers and a brief description of how they are often considered in BSM models.

\begin{table}
\begin{tabular}{@{}LLp{7.8cm}@{}}
\toprule
\text{Field} & \text{Quantum \#} & Description \\
\midrule 
N & \QuantumNumbers{1}{1}{0} & Right handed neutrino -- type I seesaw \\
E & \QuantumNumbers{1}{1}{-1} & Right handed electron \\
\Sigma & \QuantumNumbers{1}{3}{0} & Triplet right handed neutrino -- type III seesaw \\
\Sigma_1 & \QuantumNumbers{1}{3}{-1} & Triplet right handed electron \\
\mathcal{W} & \QuantumNumbers{1}{3}{0} & Triplet of bosons -- \(W^{\pm \prime}\) and \(Z'\) \\
\mathcal{L}_1 & \QuantumNumbers{1}{2}{\sfrac{1}{2}} & Doublet of bosons -- cannot contribute if they are gauge bosons from extending the SM gauge group \\
\bottomrule
\end{tabular}
\caption{New fields that can explain the CKM anomaly. The quantum numbers are given in the form \((SU(3)_c, SU(2)_L, U(1)_Y)\).}
\label{tab:BSM_models}
\end{table}

Five of these six are relatively standard ideas for BSM physics, but the final vector boson \(\mathcal{L}_1\) requires some brief further discussion.
If this vector boson is a gauge boson of an extended but spontaneously broken gauge symmetry, its renormalisable interactions with SM particles only appear in certain gauges (in particular, they vanish in the unitary gauge) and so this option might be ignored.
However, a massive \(\mathcal{L}_1\) boson can be generated in other ways, and so for completeness we consider it here but do not suggest any scenario in which it could arise.
\footnote{See also the discussion in footnote 8 of Ref.~\cite{deBlas:2017xtg}.}

Some of these models and their effect on the Cabibbo anomaly have been explored before in the literature -- the right handed neutrino \(N\) in \cite{Coutinho:2019aiy}, a combination of \(E\), \(\Sigma_1\) and two other BSM fermions in \cite{Endo:2020tkb}, the triplet vector boson \(\mathcal{W}\) in \cite{Capdevila:2020rrl}, and all four new fermion possibilities very recently in \cite{Crivellin:2020ebi} -- but we believe this work is the first to explore all the possibilities in terms of resolving the tension between EWPO and the Cabibbo anomaly.

We now move onto stage 2, studying the BSM fields and writing down all the additional SMEFT coefficients that are generated.
For each field, we write down the new Lagrangian terms and then the coefficients.
In the cases of the new vector bosons, we write \(\ldots\) to signify that there are further terms in their Lagrangians, but that these contain new coupling constants that can be set to zero without altering the generation of \(\CHlthree\).
The definitions of all the operators below can be found in \cref{app:smeft_definitions}.
For now, we assume all NP couplings are possibly complex and keep all the SM Yukawa terms such that the Lagrangians and SMEFT coefficients we show are completely general, but we will make some simplifying assumptions later in \cref{sub:BSM_fit}.

\paragraph{\(N\):}
\begin{equation}
\Lagrangian_N \supset - (\lambda_N)_i \overline{N_R} \tilde{H}^\dagger L_i + \hc \,,
\end{equation}
which generates
\begin{equation}
\label{eq:N_SMEFT_wc}
\left[C_5\right]_{ij} = \frac{(\lambda_N)_i (\lambda_N)_j}{2 M_N} \,, \quad \CHlthreebracket_{ij} = -\left[\CHlone\right]_{ij} = -\frac{(\lambda_N^*)_i (\lambda_N)_j}{4 M_N^2}
\end{equation}
after being integrated out.
\(C_5\) is the coefficient of the Weinberg operator, which generates a Majorana mass for the SM neutrinos.

\paragraph{\(E\):}
\begin{equation}
\Lagrangian_E \supset - (\lambda_E)_i \overline{E_R} H^\dagger L_i + \hc \,,
\end{equation}
which generates
\begin{equation}
\label{eq:E_SMEFT_wc}
\left[C_{eH}\right]_{ij} = \frac{(Y_e^*)_{jk} (\lambda_E^*)_i (\lambda_E)_k}{2 M_E^2} \,,  \quad \CHlthreebracket_{ij} = \left[\CHlone\right]_{ij} = -\frac{(\lambda_E^*)_i (\lambda_E)_j}{4 M_E^2}
\end{equation}
after being integrated out.

\paragraph{\(\Sigma\):}
\begin{equation}
\Lagrangian_\Sigma \supset - \frac{1}{2} (\lambda_\Sigma)_i \overline{\Sigma^a_R} \tilde{H}^\dagger \sigma^a L_i + \hc \,,
\end{equation}
which generates
\begin{equation}
\label{eq:Sigma_SMEFT_wc}
\begin{aligned}
\left[C_5\right]_{ij} = \frac{(\lambda_\Sigma)_i (\lambda_\Sigma)_j}{8 M_\Sigma} \,, \quad \left[C_{eH}\right]_{ij} = \frac{(Y_e^*)_{jk} (\lambda_\Sigma^*)_i (\lambda_\Sigma)_k}{4 M_\Sigma^2} \,, \CHlthreebracket_{ij} = \frac{1}{3} \left[\CHlone\right]_{ij} = \frac{(\lambda_\Sigma^*)_i (\lambda_\Sigma)_j}{16 M_\Sigma^2}
\end{aligned}
\end{equation}
after being integrated out.

\paragraph{\(\Sigma_1\):}
\begin{equation}
\Lagrangian_{\Sigma_1} \supset - \frac{1}{2} (\lambda_{\Sigma_1})_i \overline{\Sigma^a_{1R}} H^\dagger \sigma^a L_i + \hc \,,
\end{equation}
which generates
\begin{equation}
\label{eq:Sigma1_SMEFT_wc}
\left[C_{eH}\right]_{ij} = \frac{(Y_e^*)_{jk} (\lambda_{\Sigma_1}^*)_i (\lambda_{\Sigma_1})_k}{8 M_{\Sigma_1}^2} \,, \quad \CHlthreebracket_{ij} = -\frac{1}{3} \left[\CHlone\right]_{ij} = \frac{(\lambda_{\Sigma_1}^*)_i (\lambda_{\Sigma_1})_j}{16 M_{\Sigma_1}^2}
\end{equation}
after being integrated out.

\paragraph{\(\mathcal{W}\):}
\begin{equation}
\Lagrangian_{\mathcal{W}} \supset - \frac{1}{2} (\lambda_\mathcal{W}^L)_{ij} \overline{L_i} \sigma^a \gamma^\mu L_j \mathcal{W}^a_\mu - \left( \frac{i}{2} \lambda_\mathcal{W}^H \mathcal{W}^a_\mu H^\dagger \sigma^a D_\mu H + \hc \right) + \ldots \,,
\end{equation}
which generates
\begin{equation}
\begin{aligned}
\CHlthreebracket_{ij} &= - \frac{\Re(\lambda_\mathcal{W}^H) (\lambda_\mathcal{W}^L)_{ij}}{4 M_\mathcal{W}^2} \,, \quad \CHlone = 0 \,, \quad \left[C_{\ell\ell}\right]_{ijkl} = \frac{(\lambda_\mathcal{W}^L)_{ij} (\lambda_\mathcal{W}^L)_{kl} - 2 (\lambda_\mathcal{W}^L)_{il} (\lambda_\mathcal{W}^L)_{kj}}{8 M_\mathcal{W}^2} \,, \\
\left[C_{eH}\right]_{ij} &= \frac{i (Y_e^*)_{jk} (\lambda_\mathcal{W}^L)_{ik} \Im(\lambda_\mathcal{W}^H)}{4 M_\mathcal{W}^2} - \frac{i(Y_e^\dagger)_{ij} \Im\left[(\lambda_\mathcal{W}^H)^2\right]}{8 M_\mathcal{W}^2} \,, \\
\left[C_{dH}\right]_{ij} &= - \frac{i(Y_d^\dagger)_{ij} \Im\left[(\lambda_\mathcal{W}^H)^2\right]}{8 M_\mathcal{W}^2} \,, \quad \left[C_{uH}\right]_{ij} = - \frac{i(Y_u^\dagger)_{ij} \Im\left[(\lambda_\mathcal{W}^H)^2\right]}{8 M_\mathcal{W}^2} \,, \\
C_H &= -\frac{\lambda |\lambda_\mathcal{W}^H|^2}{M_\mathcal{W}^2} + \frac{\mu^2 |\lambda_\mathcal{W}^H|^4}{2 M_\mathcal{W}^4} \,, \quad C_{HD} = \frac{|\lambda_\mathcal{W}^H|^2 - \Re\left[(\lambda_\mathcal{W}^H)^2\right]}{4 M_\mathcal{W}^2} \,, \quad C_{H\square} = - \frac{\Re\left[(\lambda_\mathcal{W}^H)^2\right]}{8 M_\mathcal{W}^2}
\end{aligned}
\end{equation}
after being integrated out.
Here \(\mu, \lambda\) are the SM coefficients of the Higgs doublet and quartic terms respectively.

\paragraph{\(\mathcal{L}_1\):}
\begin{equation}
\Lagrangian_{\mathcal{L}_1} \supset - \left( \gamma_{\mathcal{L}_1} \mathcal{L}_1^{\mu \dagger} D_\mu H + \hc \right) - i \lambda_{\mathcal{L}_1}^W \mathcal{L}_1^{\mu \dagger} \sigma^a \mathcal{L}_1^\nu W^a_{\mu \nu} + \ldots \,,
\end{equation}
which generates
\begin{equation}
\begin{aligned}
\CHlthreebracket_{ij} &= \delta_{ij} \frac{g \lambda_{\mathcal{L}_1}^W |\gamma_{\mathcal{L}_1}|^2}{4 M_{\mathcal{L}_1}^4 Z_H} \,, \quad \left[C_{Hq}^{(3)}\right]_{ij} = \delta_{ij} \frac{g \lambda_{\mathcal{L}_1}^W |\gamma_{\mathcal{L}_1}|^2}{4 M_{\mathcal{L}_1}^4 Z_H} \,, \\
\quad C_H &= \frac{2 g \lambda_{\mathcal{L}_1}^W \lambda |\gamma_{\mathcal{L}_1}|^2}{M_{\mathcal{L}_1}^4 Z_H} \,, \quad C_{H\square} = \frac{3 g \lambda_{\mathcal{L}_1}^W |\gamma_{\mathcal{L}_1}|^2}{4 M_{\mathcal{L}_1}^4 Z_H^2} \,, \\
C_{HB} &= -\frac{g^{\prime 2} |\gamma_{\mathcal{L}_1}|^2}{8 M_{\mathcal{L}_1}^4 Z_H} \,, \quad C_{HW} = -\frac{g (g + 2 \lambda_{\mathcal{L}_1}^W) |\gamma_{\mathcal{L}_1}|^2}{8 M_{\mathcal{L}_1}^4 Z_H} \,, \quad C_{HWB} = -\frac{g' (g + \lambda_{\mathcal{L}_1}^W) |\gamma_{\mathcal{L}_1}|^2}{4 M_{\mathcal{L}_1}^4 Z_H} \,, \\
\left[C_{eH}\right]_{ij} &= \frac{g \lambda_{\mathcal{L}_1}^W |\gamma_{\mathcal{L}_1}|^2 (Y_e^\dagger)_{ij}}{2 M_{\mathcal{L}_1}^4 Z_H} \,, \quad \left[C_{dH}\right]_{ij} = \frac{g \lambda_{\mathcal{L}_1}^W |\gamma_{\mathcal{L}_1}|^2 (Y_d^\dagger)_{ij}}{2 M_{\mathcal{L}_1}^4 Z_H} \,, \quad \left[C_{uH}\right]_{ij} = \frac{g \lambda_{\mathcal{L}_1}^W |\gamma_{\mathcal{L}_1}|^2 (Y_u^\dagger)_{ij}}{2 M_{\mathcal{L}_1}^4 Z_H}
\end{aligned}
\end{equation}
after being integrated out.
Here
\begin{equation}
Z_H = 1 - \frac{|\gamma_{\mathcal{L}_1}|^2}{M_{\mathcal{L}_1}^2}
\end{equation}
reflects the different normalisation of the Higgs doublet in a BSM model with the \(\mathcal{L}_1\) vector.

\subsection{BSM fits}
\label{sub:BSM_fit}

We now complete the third part of our process outlined at the beginning of this section.
This is broken up into two parts -- first we make a BSM fit for the fermions, where the relationship between the singlet and triplet operators \(Q_{H\ell}^{(1,3)}\) are fixed in the way required by each BSM scenario, since each fermion essentially only generates these operators.
The idea of this exercise is to examine how new physics in the singlet \(\CHlone\), which affects only \PZ couplings and is therefore well placed to change the results of the EWPO fit, could reduce the tension.
Secondly, we do specific fits for the two vector bosons as they generate a wider and distinct range of operators.
For the purpose of these fits, we make two simplifying assumptions that were alluded to earlier -- that all the new couplings are real, and neglecting all SM Yukawas except that of the top quark.

\subsubsection{BSM fermions}
\begin{figure}
\includegraphics[width=0.65\textwidth]{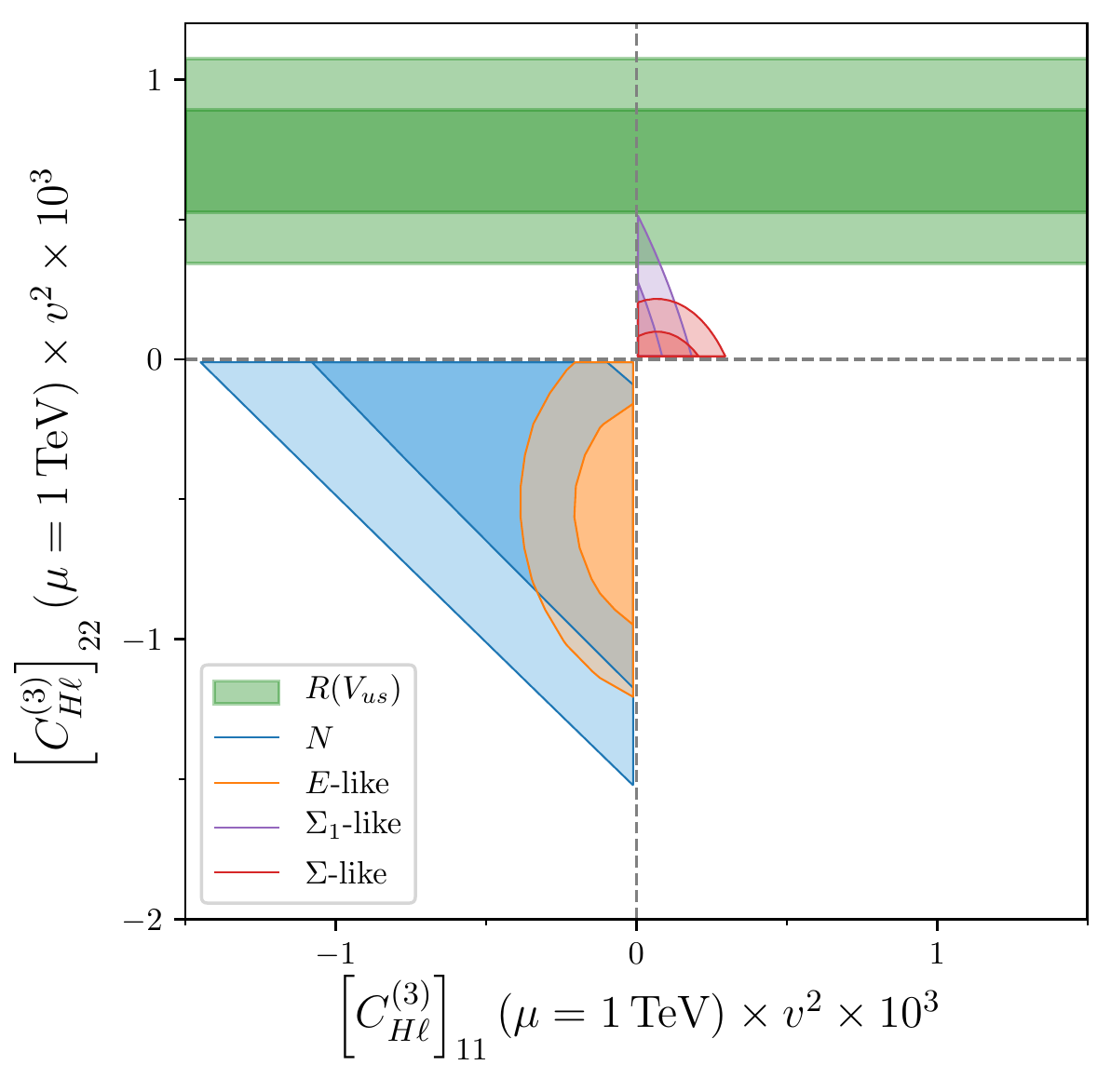}
\caption{Fits to EWPO in BSM fermion scenarios for each of the models we consider -- \(N\) (blue), \(E\) (yellow), \(\Sigma\) (red), and \(\Sigma_1\) (purple). See the main text for the caveats such that the latter three regions are ``BSM-like''.}
\label{fig:BSM-like}
\end{figure}

The fit results for the fermions is shown in \cref{fig:BSM-like}.
Before discussing what this figure tells us, we note two pieces of information.
Firstly, that each allowed region lives in a single quadrant of the figure, which is caused by the form of the generated coefficients -- examining \cref{eq:N_SMEFT_wc,eq:E_SMEFT_wc,eq:Sigma_SMEFT_wc,eq:Sigma1_SMEFT_wc} we see that \(\CHlthreebracket_{ii} \sim \pm |\lambda_i|^2 / M^2\) and hence they have a fixed sign.
As such, there is no way for the new lepton states \(N\) or \(E\) alone to generate correct sign to explain the \RVus anomaly.
Secondly, since the fermions have only a single new Yukawa-like coupling to the SM, off-diagonal elements of the SMEFT coefficients are inevitable if they couple to more than one generation -- schematically we have \(\CHlthreebracket_{ij} \sim \sqrt{\CHlthreebracket_{ii} \CHlthreebracket_{jj}}\).
Such off-diagonal elements are very strongly constrained by measurements of lepton flavor violating (LFV) effects, which are generated at tree level by all our fermions except the right handed neutrino.
The constraints are so strong that we can effectively say that, with the exception of the right handed neutrino \(N\), the allowed regions really must lie on either of the axes, as indicated in the figure by the grey dashed lines, to avoid the experimental constraints on LFV.

With these in mind we now see that, in terms of a single particle extension, the triplet \(\Sigma_1\) with couplings only to the second generation is best placed to generate a scenario with agreement between \RVus and EWPO.
\footnote{One might wonder whether such a new particle with sizeable couplings to muons could help explain the muon anomalous magnetic moment, since currently the SM prediction \cite{Aoyama:2020ynm} is \SI{3.7}{\sigma} below the experimental result \cite{Bennett:2006fi}.
Unfortunately, the contribution of the \(\Sigma_1\) is around two orders of magnitude smaller than the discrepancy, for the size of coupling over mass implied here, and anyway is of the wrong sign.}
We observe however that the best fit for \RVus in this case is still in tension with EWPO at the level of more than \SI{2}{\sigma}, and so does not provide a sizeable improvement over the simple situation with \(\CHlthree\) alone.
In fact, the allowed region for a new \(\Sigma_1\) field is effectively centred on the origin (as is the \(\Sigma\) region as well), which as we will see in \cref{sec:future} will become a problem in the future as experimental precision on the EWPO increases.
Conversely the \(N\) and \(E\) fields, for which there is perhaps some small evidence of a non-zero effect, are unable to generate the correct sign to match \RVus as has already been discussed.

\subsubsection{BSM vector bosons}
\begin{figure}
\includegraphics[width=0.65\textwidth]{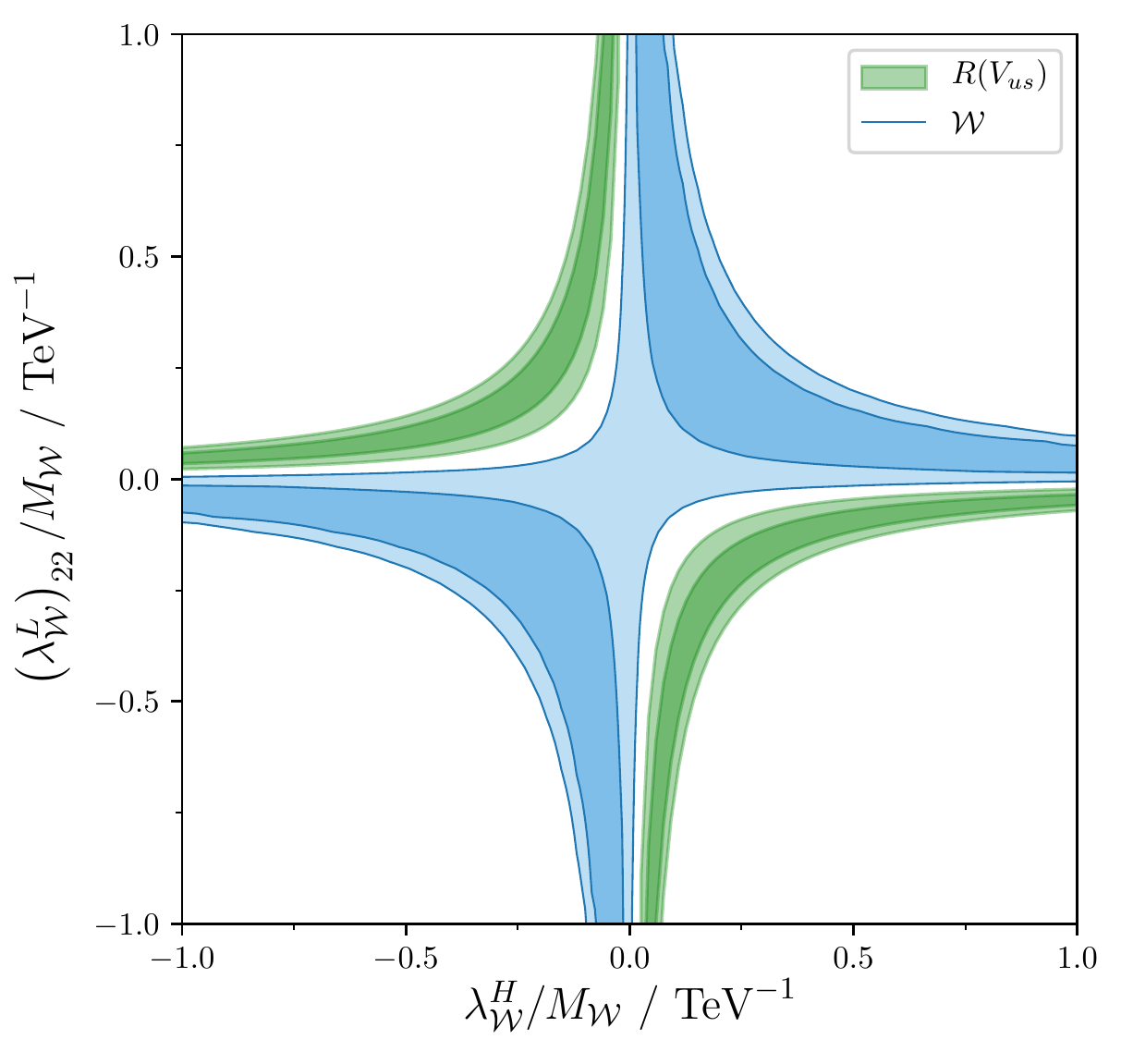}
\caption{Fit to EWPO in the scenario of the new vector boson \(\mathcal{W}\).}
\label{fig:BSMWvector}
\end{figure}

\begin{figure}
\includegraphics[width=0.49\textwidth]{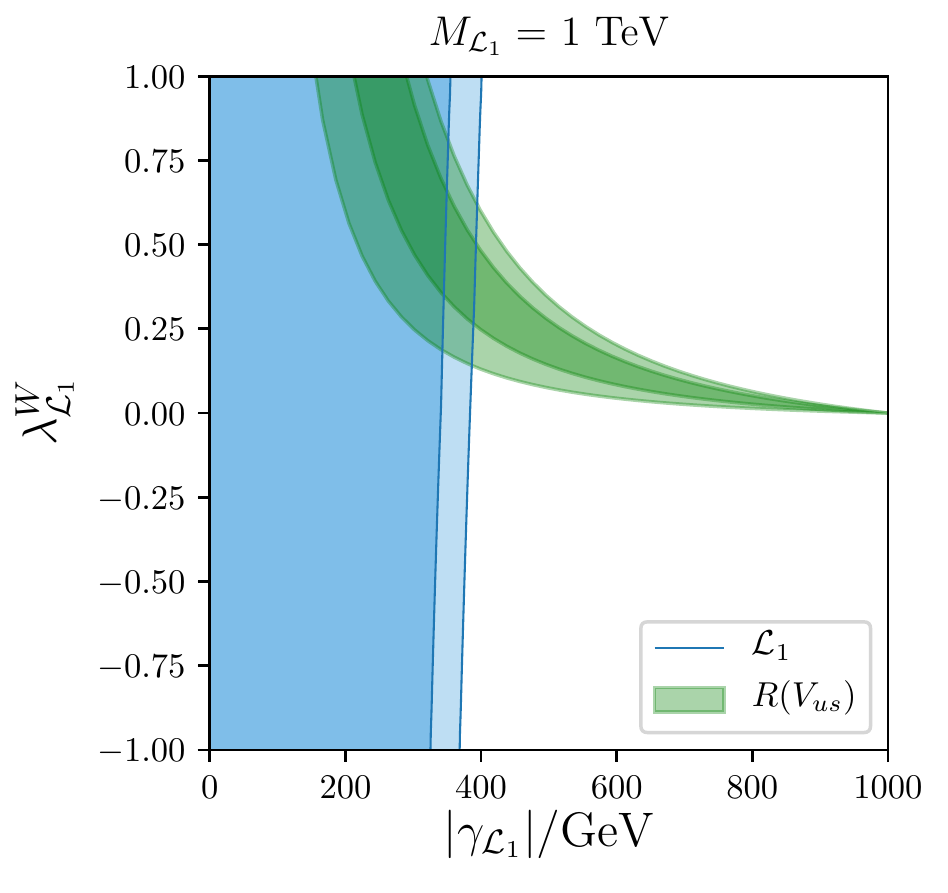}
\includegraphics[width=0.49\textwidth]{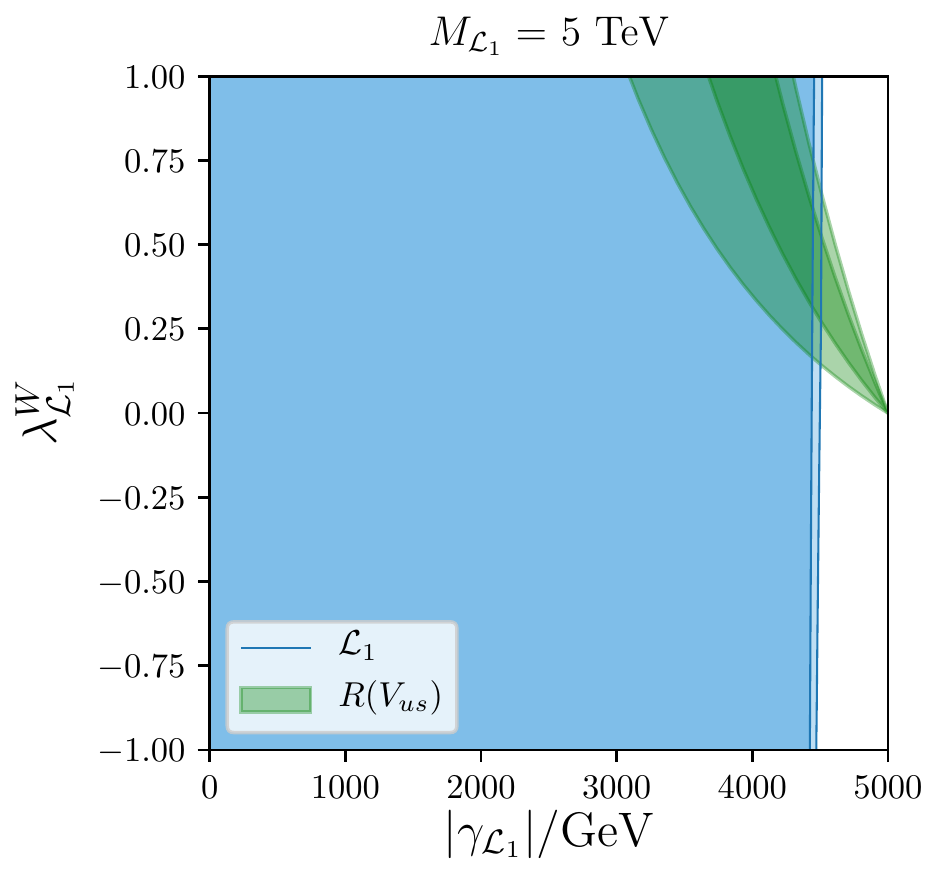}
\caption{Fit to EWPO in the scenario of the new vector boson \(\mathcal{L}_1\) for two different benchmark masses of \SI{1}{\TeV} (left) and \SI{5}{\TeV} (right).}
\label{fig:BSML1vector}
\end{figure}

The fit results for the two vector boson extensions \(\mathcal{W}\) and \(\mathcal{L}_1\) are shown in \cref{fig:BSMWvector,fig:BSML1vector} respectively.
For the triplet of bosons \(\mathcal{W}\) we can easily parameterise all the SMEFT contributions in terms of the two ratios \(\lambda_\mathcal{W}^H / M_\mathcal{W}\) and \((\lambda_\mathcal{W}^L)_{22} / M_\mathcal{W}\) and we see from the figure that the tension is unresolved.
(Later, in \cref{sub:resolve_tension}, we examine a future scenario where this tension is reduced and so this new boson could become more plausible.)
This is not unexpected, since the singlet coefficient \(\CHlone\) is exactly zero in this model and so most EWPO are modified in exactly the same way as the simple situation examined in \cref{sec:global_fit}.

The \(\mathcal{L}_1\) extension is more complicated, as it is not possible to parameterise the SMEFT contributions in terms of two ratios.
Being specific, the three coefficients \(C_{HB}\), \(C_{HW}\), and \(C_{HWB}\) depend on the couplings \(\gamma_{\mathcal{L}_1}\) and \(\lambda_{\mathcal{L}_1}^W\) and the vector mass independently.
We therefore examine two benchmark scenarios, where the new vector boson mass is fixed to \SI{1}{\TeV} and \SI{5}{\TeV}, which are shown in \cref{fig:BSML1vector}.
At both benchmark points, we see that there is parameter space where this new boson can explain both the Cabibbo anomaly and current EWPO data simultaneously.
For the light benchmark, the coupling \(\lambda_{\mathcal{L}_1}^W\) to the \(SU(2)_L\) field strength tensor can be small with a dimensionful coupling to the Higgs near the electroweak scale, while at the heavy benchmark perturbative values of \(\lambda_{\mathcal{L}_1}^W\) can only explain the data with a much larger value of \(\gamma_{\mathcal{L}_1}\) at the multi-TeV scale.
Since, as we have discussed earlier the \(\mathcal{L}_1\) cannot contribute if it arises through extending the gauge symmetry of the SM, and more generally in any complete unitary theory extending the SM, we consider that a realistic BSM scenario of this type is likely to be hard to construct.
As such, we now move on and leave this possibility as an area for detailed future study.

\section{Future of EWPO}
\label{sec:future}

In the previous sections we established that the current status of EWPO means that there is a large (greater than \SI{3}{\sigma}) tension between those measurements and the new \RVus observable, assuming that the the triplet operator \(Q_{H\ell}^{(3)}\) is the ``source'' of the \RVus discrepancy, and that looking at realistic NP extensions to the Standard Model does not provide a clear way to reduce the tension.
As such, we now look in more detail at the electroweak fit and what is driving the tension, and at how these EWPO measurements could change in the future.

\subsection{Details of the EW fit}

\begin{figure}
\includegraphics[width=0.9\textwidth]{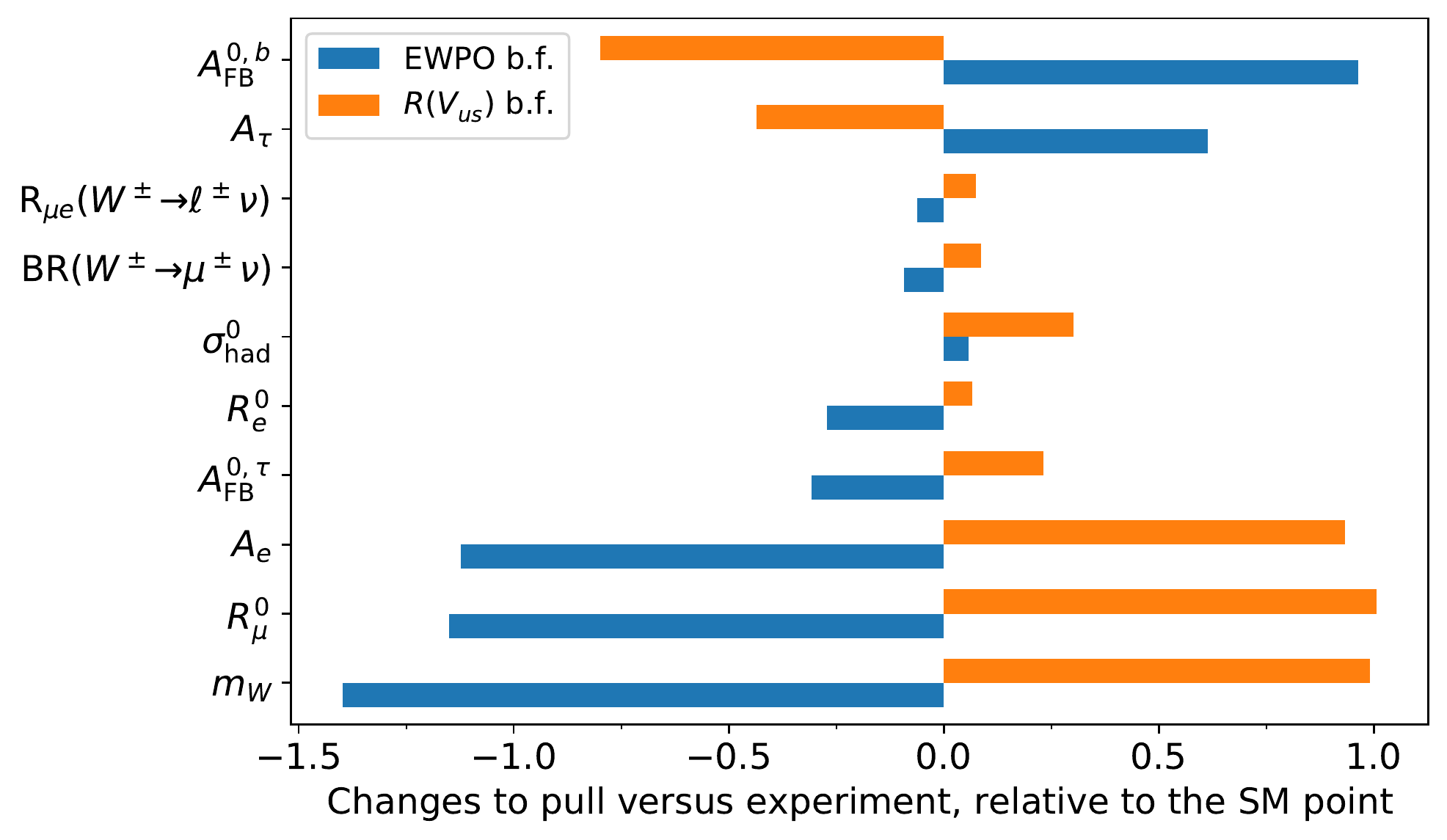}
\caption{A visual representation of the important changes to the pulls of EWPO observables at the EWPO best fit from \cref{eq:EWPO_bf} (blue) and \RVus best fit from \cref{eq:my_CHl322_number} (orange), relative to the SM.}
\label{fig:EWPO_pull_changes}
\end{figure}

If we delve inside the EW fit, we find that there are 10 observables
\footnote{We remind the reader than \cref{tab:smelli_obs_EWPT} contains the descriptions of the electroweak precision observables under consideration.}
where the difference between the current experimental average and the SM prediction is greater than one standard deviation, which can be seen in \cref{tab:ewpo_pulls} in the `SM' column.
Of these, one (\(\text{BR}(W \to \tau \nu)\)) has been re-measured by ATLAS very recently \cite{Aad:2020ayz} and found to be much closer to the SM than the old LEP result \cite{Schael:2013ita}.
Another five show very little change when the theoretical prediction is evaluated at the \RVus best fit point (corresponding to \cref{eq:my_CHl322_number}) or at the best fit of the EW fit in \cref{fig:global_and_EWPT_vs_RVus} (\cref{eq:EWPO_bf}), and so are not sensitive enough to the NP effects under discussion.
This leaves us with four sensitive observables that can be considered to be driving the fit: \(R_\mu^0, m_W, A_e, A_\text{FB}^{0,b}\).
The first three improve with respect to experiment at the EWPO best fit point, with each showing a reduction in pull of more than \SI{1}{\sigma}, while only \(A_\text{FB}^{0,b}\) becomes yet more discrepant from the measurement but by a smaller amount, which demonstrates why the fit shows the result it does.
A graphical representation of these changes can be seen in \cref{fig:EWPO_pull_changes}, where for clarity we have only shown those observables which have a pull greater than \SI{1}{\sigma} between the SM and experiment or change by at least \SI{1}{\sigma} between the EWPO and \RVus best fit.
These four observables are therefore the ideal ones in which a future change in value and/or precision could significantly affect the fits as described so far.

\begin{table}
\begin{tabular}{@{}lCCC@{}}
\toprule
Observables & \text{SM} & \text{EWPO best fit} & \text{\RVus best fit} \\
\midrule 
$A_\mathrm{FB}^{0, b}$                        &  2.4 &  3.4 &  1.6 \\
$A_\tau$                                      &  0.9 &  1.5 &  0.5 \\
$A_\mathrm{FB}^{0, c}$                        &  0.8 &  1.2 &  0.6 \\
$A_\mathrm{FB}^{0, e}$                        &  0.7 &  0.9 &  0.5 \\
$R_\tau^0$                                    &  0.4 &  0.5 &  0.3 \\
$A_\mu$                                       &  0.3 &  0.4 &  0.3 \\
$\Gamma_W$                                    &  0.2 &  0.2 &  0.1 \\
$R_b^0$                                       &  0.7 &  0.7 &  0.7 \\
$A_b$                                         &  0.6 &  0.6 &  0.6 \\
$R_c^0$                                       &  0.0 &  0.1 &  0.0 \\
$\mathrm{R}_{\tau  e}(W^\pm \to \ell^\pm\nu)$ &  0.7 &  0.7 &  0.7 \\
$A_s$                                         &  0.5 &  0.5 &  0.5 \\
$R_{uc}^0$                                    &  0.7 &  0.7 &  0.7 \\
$\mathrm{BR}(W^\pm \to  e^\pm\nu)$            &  0.8 &  0.8 &  0.8 \\
$\mathrm{R}(W^+\to cX)$                       &  0.3 &  0.3 &  0.3 \\
$\mathrm{BR}(W^\pm \to \tau^\pm\nu)$          &  2.6 &  2.6 &  2.6 \\
$A_c$                                         &  0.1 &  0.1 &  0.1 \\
$\mathrm{R}_{\mu  e}(W^\pm \to \ell^\pm\nu)$  &  1.1 &  1.0 &  1.2 \\
$\mathrm{BR}(W^\pm\to \mu^\pm\nu)$            &  1.4 &  1.3 &  1.5 \\
$\sigma_\mathrm{had}^0$                       &  1.5 &  1.6 &  1.8 \\
$R_e^0$                                       &  1.4 &  1.1 &  1.5 \\
$A_\mathrm{FB}^{0,\mu}$                       &  0.5 &  0.3 &  0.7 \\
$A_\mathrm{FB}^{0,\tau}$                      &  1.5 &  1.2 &  1.8 \\
$\Gamma_Z$                                    &  0.5 &  0.7 &  1.3 \\
$A_e$                                         &  2.2 &  1.1 &  3.2 \\
$R_\mu^0$                                     &  1.5 &  0.4 &  2.6 \\
$m_W$                                         &  1.7 &  0.3 &  2.7 \\
\bottomrule
\end{tabular}
\caption{The pulls (measured in sigmas) of different observables within the EW fit relative to experiment, at the SM point where all SMEFT coefficients are zero, at the best fit of the EWPO only fit (\cref{eq:EWPO_bf}), and at the \RVus best fit (\cref{eq:my_CHl322_number}). They are ordered by the difference between the current EWPO best fit and the \RVus best fit.}
\label{tab:ewpo_pulls}
\end{table}

\subsection{Future measurements}

We consider two possibilities in terms of future measurements and increases in precision -- a ``near-future'' case similar to the ILC, for which we take projections from \cite{Baer:2013cma}, and a ``far-future'' experiment like the FCC-ee where we use projections from \cite{Abada:2019zxq,FCC-ee:2019talk}.
The ILC TDR quotes several specific numerical predictions for improvements to the EW observables in Table~4.10 of \cite{Baer:2013cma}, while we take numerics for the FCC-ee from Sections~1.2.2-4 of \cite{Abada:2019zxq} plus a prediction for the \PW leptonic decay with from Slide 22 of \cite{FCC-ee:2019talk} -- these are summarised in \cref{tab:future_numerics} along with the current experimental uncertainties.

\begin{table}
\begin{tabular}{@{}LLLL@{}}
\toprule
\text{Observable} & \text{Current} & \text{ILC} & \text{FCC-ee} \\
\midrule 
m_W & \SI{\pm 12}{\MeV} & \SI{\pm 6}{\MeV} & \SI{\pm 0.5}{\MeV} \\
\Gamma_Z & \SI{\pm 2.3}{\MeV} & \SI{\pm 0.8}{\MeV} & \SI{\pm 0.12}{\MeV} \\
A_b & \num{\pm 20e-3} & \num{\pm 1e-3} & \multicolumn{1}{c}{$\cdots$} \\
R^0_b & \num{\pm 6.6e-2} & \num{\pm 1.4e-2}{} & \multicolumn{1}{c}{$\cdots$} \\
\Gamma_W & \SI{\pm 42}{\MeV} & \multicolumn{1}{c}{$\cdots$} & \SI{\pm 1}{\MeV} \\
R^0_e & \num{\pm 50e-3} & \multicolumn{1}{c}{$\cdots$} & \num{\pm 1e-3} \\
R^0_\mu & \num{\pm 33e-3} & \multicolumn{1}{c}{$\cdots$} & \num{\pm 1e-3 } \\
R^0_\tau & \num{\pm 45e-3} & \multicolumn{1}{c}{$\cdots$} & \num{\pm 1e-3 } \\
R_{\mu e} (W \to \ell \nu) & \num{\pm 8} & \multicolumn{1}{c}{$\cdots$} & \num{\pm 4 e-2} \\
\bottomrule
\end{tabular}
\caption{Future improvements to precision of EW observables, in the various scenarios considered. The current experimental uncertainties are from the 2020 PDG \cite{Zyla:2020zbs}.}
\label{tab:future_numerics}
\end{table}

\subsection{Resolving the tension?}
\label{sub:resolve_tension}
\begin{figure}
\includegraphics[width=0.65\textwidth]{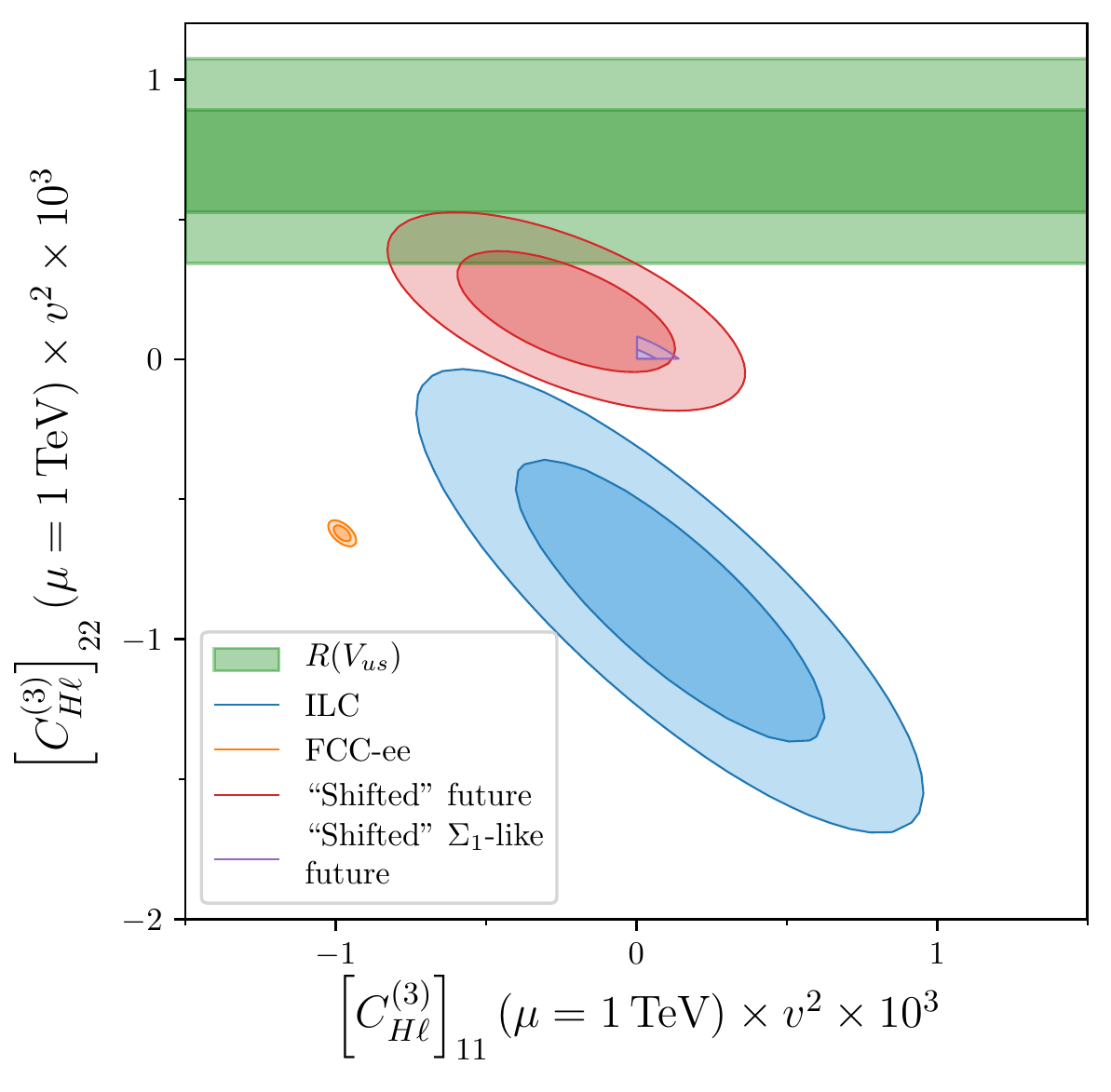}
\caption{Fits to EWPO in our future scenarios that are ``near-future'' (blue), ``far-future'' (orange), and ``shifted''(red), all assuming only NP in \(\CHlthree\), as well as the possibility of the \(\Sigma_1\) scenario in the ``shifted'' future (purple).}
\label{fig:all_futures_vs_RVus}
\end{figure}

With both the possible future precision improvements in mind as well as knowing which observables are driving the fit, we now examine some future scenarios and what the electroweak fit looks like therein.
Two of our scenarios are basic extrapolations using the same central values as today, simply applying the improvement in precision expected from the ILC and FCC-ee machines.
These are shown in \cref{fig:all_futures_vs_RVus} in blue and orange respectively.
The ``near-future'' scenario replicated the general pattern of the current status (i.e. zero \(\CHlthreebracket_{11}\), negative \(\CHlthreebracket_{22}\)) with an increased significance -- numerically the blue oval corresponds to
\begin{equation}
\begin{aligned}
\CHlthreebracket_{11} &= \SI{0.18 \pm 0.56 e-8}{\GeV^{-2}} \,, \\
\CHlthreebracket_{22} &= \SI{-1.4 \pm 0.56 e-8}{\GeV^{-2}} \,,
\end{aligned}
\end{equation}
with a strong correlation of \(-0.83\).
In this case, and with no change to \RVus, the tension increases from the current \SI{3.6}{\sigma} to \SI{4}{\sigma}.
A full set of data from a ``far-future'' machine brings with it a revolution in precision -- it is clear that the small orange oval does not agree in any way with the deviation in \RVus and so in such a future there must be some other mechanism at work.

Our final scenario is more interesting, and involves considering if the observables driving the current EW fit shift are re-measured with different central values
such that EWPO point in the same direction as \RVus -- we call this the ``shifted'' scenario.
We imagine that \(m_W\), \(R_\mu^0\), and \(A_e\) have their central values shifted (downwards) by twice the current experimental uncertainties.
\footnote{Since the EW sub fit contains 27 separate observables, such a change in three alone is a reasonably plausible scenario.}
On top of this shift, we use the improved precision of the ``near-future'' scenario.
\footnote{For \(R_\mu^0\) we assume the same relative improvement from now as for \(R_b^0\), while for \(A_e\) we use the same absolute precision as \(A_b\), which are conservative choices.}
This scenario is shown in red in \cref{fig:all_futures_vs_RVus}. 
We see that such an imagined scenario would give a clear improvement in the internal tension of a global fit and could point towards a consistent BSM effect at work.
However when we examine the particular case of the \(\Sigma_1\) field that, as we saw earlier in \cref{sub:BSM_fit}, came the closest to reconciling the current discrepancy, we see a problem.
As noted earlier, current data (see the purple region in \cref{fig:BSM-like}) seems to favour no new physics with this pattern of SMEFT coefficients and the increased precision of the ILC (or something similar) merely reduces the size of the \SI{2}{\sigma} allowed region, despite our hypothesis of the most discrepant observables being measured as closer to the predictions required by the fitted value of \RVus.
As such, the \(SU(2)\) triplet field would become a much less plausible solution unless there is even more significant change to observations of EWPO than that which we have investigated.

\section{Conclusions}
\label{sec:summary}

In this article we considered the newly emerging Cabibbo anomaly and have showed how the recent improvements in radiative corrections mean the anomaly is now at the \SI{4}{\sigma} level.
After summarising how a recent work \cite{Crivellin:2020lzu} had made an argument for a new observable \RVus and new physics in the leptonic \PW vertex, we expand their simple modification of the SM by analysing it from the perspective of the SMEFT.
We found that the particular low energy change to the \PW vertex is exclusively generated the by operator \(Q_{H\ell}^{(3)}\) once you impose the requirement of allowing a LFUV effect, as described in \cref{sec:SMEFT}.
The Cabibbo anomaly then corresponds to a non-zero value in the coefficient \(\CHlthreebracket_{22}\) which we gave in \cref{eq:my_CHl322_number}.

Having identified that operator, we performed a global fit to a large number of observables using the \smelli software package and found a tension exists between that global fit and the \RVus observable that has a high sensitivity to the assumed pattern of new physics couplings.
Looking deeper, we discovered that the global fit is dominated by electroweak precision observables, and that singling those out there was in fact a large tension between EWPO and the Cabibbo anomaly favoured region of parameter space, as seen clearly in \cref{fig:global_and_EWPT_vs_RVus}, at the level of \SI{3.6}{\sigma}.
In light of this tension, we used the results of \cite{deBlas:2017xtg} to systematically identify all the possible single extensions of the SM that could generate the operator of interest (which amounted to four fermions and two vector bosons) and then documented all the other correlated SMEFT coefficients that are also induced in those specific BSM models, which are detailed in \cref{sec:BSM}.
After performing a fit to the electroweak precision data again within each BSM scenario, we find that no unitary extension of the SM by a single new particle can cause a significant reduction in the tension we identified, but that a heavy \(SU(2)_L\) triplet \(\Sigma_1\) that couples exclusively to second generation leptons provides the largest reduction.

In light of this result, we then proceeded in \cref{sec:future} to analyse the future of EWPO given predicted improvements at the ILC and FCC-ee future experiments.
By looking at the individual observables in the EWPO subset, we identified a small number that drive the current tension (see for example \cref{fig:EWPO_pull_changes}) that are also set to be re-measured at a higher level of precision at future colliders.
We found that, if the central values are unchanged after an ILC-like machine, then the current tension between \RVus and EWPO increases to just over \SI{4}{\sigma}, even assuming no change to the Cabibbo anomaly.
However, if we forecast a future where three specific observables (\(M_W, R_\mu^0, A_e\)) out of the full set have changed by two standard deviations each, we could end up with almost complete agreement between all data in our simplified scenario where only a single SMEFT operator \(\CHlthree\) is active (which could be achieved by the new vector boson \(\mathcal{W}\) coupling exclusively to the second generation lepton doublets).
Examining a more realistic pattern of coefficients, like that generated by the \(SU(2)_L\) triplet field \(\Sigma_1\), the increased precision from a near future machine almost eliminates the possibility of it providing a combined explanation of all data, without a very large shift in the observed data occurring in the future.
We must conclude that either there is a error in the current EWPO data at least of the size considered in \cref{sub:resolve_tension}, or that the hypothesis of \cite{Crivellin:2020lzu} (that the Cabbibo anomaly can be explained solely by LFUV in the leptonic \PW decay) is not consistent with any unitary single new particle scenario above the electroweak scale. 

\paragraph*{Note:} While this work was in preparation, the article \cite{Crivellin:2020ebi} appeared on the arXiv covering a similar area. While they do not examine the internals of the global fit, they also find that a \(\Sigma_1\) triplet fermion with muonic couplings gives the best fit to all current data, in agreement with one of our conclusions from \cref{sub:BSM_fit}.

\begin{acknowledgments}
I was financially supported by MIUR (Italy) under a contract PRIN 2015P5SBHT and by INFN Sezione di Roma La Sapienza and partially supported by the ERC-2010 DaMESyFla Grant Agreement Number: 267985.
I would like to thank Daniele Barducci and Darren Scott for helpful discussions at various times during the preparation of this work, and Alex Lenz and Aleksey Rusov for a helpful discussion afterwards.
\end{acknowledgments}

\appendix

\section{SMEFT definitions}
\label{app:smeft_definitions}

The SMEFT operators we use in this work are defined as follows:
\begin{align}
\left[Q_{H\ell}^{(3)}\right]_{ij} &= \left( \PH^\dagger i \overset{\leftrightarrow}{D^a_\mu} \PH \right) \left( L_i \sigma^a \gamma_\mu L_j \right) \\
\left[Q_{H\ell}^{(1)}\right]_{ij} &= \left( \PH^\dagger i \overset{\leftrightarrow}{D_\mu} \PH \right) \left( L_i \gamma_\mu L_j \right) \\
\left[Q_5\right]_{ij} &= \overline{L^c_i} \tilde{H}^* \tilde{H}^\dagger L_j \\
\left[Q_{eH}\right]_{ij} &= (H^\dagger H) (\overline{L_i} H e_j) \\
\left[Q_{dH}\right]_{ij} &= (H^\dagger H) (\overline{Q_i} H d_j) \\
\left[Q_{uH}\right]_{ij} &= (H^\dagger H) (\overline{Q_i} \tilde{H} u_j) \\
\left[Q_{\ell\ell}\right]_{ijkl} &= (\overline{L}_i \gamma^\mu L_j) (\overline{L}_k \gamma_\mu L_l) \\
Q_{H} &= (H^\dagger H)^3 \\
Q_{HD} &= \left( \PH^\dagger D^\mu \PH \right)^* \left( \PH^\dagger D_\mu \PH \right) \\
Q_{H\square} &= (H^\dagger H) \square (H^\dagger H) \\
Q_{HB} &= \left( \PH^\dagger \PH \right) B^{\mu\nu} B_{\mu\nu} \\
Q_{HW} &= \left( \PH^\dagger \PH \right) W^{a, \mu\nu} W_{\mu\nu}^a \\
Q_{HWB} &= \left( \PH^\dagger \sigma^a \PH \right) W_{\mu\nu}^a B^{\mu\nu}
\end{align}
where \(\overset{\leftrightarrow}{D^a_\mu} = \left( \sigma^a D_\mu - \overset{\leftarrow}{D}_\mu \sigma^a \right)\), \(\sigma^a\) are the Pauli matrices, and \(i,j,k,l\) are flavour generation indices.

\section{\smelli electroweak precision observables}
\label{app:ewpt_obs}

\begin{table*}[h]
\begin{tabularx}{\textwidth}{@{}LXll@{}}
\toprule
\text{Observable} & Description & Exp. & Theory \\
\midrule
\Gamma_Z & Total width of the $Z^0$ boson & \cite{ALEPH:2005ab} & \cite{Brivio:2017vri,Freitas:2014hra} \\ 
\sigma_\text{had}^0 & $e^+e^-\to Z^0$ hadronic pole cross-section & \cite{ALEPH:2005ab} & \cite{Brivio:2017vri,Freitas:2014hra} \\ 
R_ e^0 & Ratio of $Z^0$ partial widths to hadrons vs. $ e$ pairs & \cite{ALEPH:2005ab} & \cite{Brivio:2017vri,Freitas:2014hra} \\ 
R_\mu^0 & Ratio of $Z^0$ partial widths to hadrons vs. $\mu$ pairs & \cite{ALEPH:2005ab} & \cite{Brivio:2017vri,Freitas:2014hra} \\ 
R_\tau^0 & Ratio of $Z^0$ partial widths to hadrons vs. $\tau$ pairs & \cite{ALEPH:2005ab} & \cite{Brivio:2017vri,Freitas:2014hra} \\ 
A_\text{FB}^{0, e} & Forward-backward asymmetry in $Z^0\to  e^+ e^-$ & \cite{ALEPH:2005ab} & \cite{Brivio:2017vri} \\ 
A_\text{FB}^{0,\mu} & Forward-backward asymmetry in $Z^0\to \mu^+\mu^-$ & \cite{ALEPH:2005ab} & \cite{Brivio:2017vri} \\ 
A_\text{FB}^{0,\tau} & Forward-backward asymmetry in $Z^0\to \tau^+\tau^-$ & \cite{ALEPH:2005ab} & \cite{Brivio:2017vri} \\ 
A_ e & Asymmetry parameter in $Z^0\to  e^+ e^-$ & \cite{ALEPH:2005ab} & \cite{Brivio:2017vri} \\ 
A_\mu & Asymmetry parameter in $Z^0\to \mu^+\mu^-$ & \cite{ALEPH:2005ab} & \cite{Brivio:2017vri} \\ 
A_\tau & Asymmetry parameter in $Z^0\to \tau^+\tau^-$ & \cite{ALEPH:2005ab} & \cite{Brivio:2017vri} \\ 
R_ b^0 & Ratio of $Z^0$ partial widths to $ b$ pairs vs. all hadrons & \cite{ALEPH:2005ab} & \cite{Brivio:2017vri,Freitas:2014hra} \\ 
R_ c^0 & Ratio of $Z^0$ partial widths to $ c$ pairs vs. all hadrons & \cite{ALEPH:2005ab} & \cite{Brivio:2017vri,Freitas:2014hra} \\ 
A_\text{FB}^{0, b} & Forward-backward asymmetry in $Z^0\to b\bar b$ & \cite{ALEPH:2005ab} & \cite{Brivio:2017vri} \\ 
A_\text{FB}^{0, c} & Forward-backward asymmetry in $Z^0\to c\bar c$ & \cite{ALEPH:2005ab} & \cite{Brivio:2017vri} \\ 
A_ b & Asymmetry parameter in $Z^0\to b\bar b$ & \cite{ALEPH:2005ab} & \cite{Brivio:2017vri} \\ 
A_ c & Asymmetry parameter in $Z^0\to c\bar c$ & \cite{ALEPH:2005ab} & \cite{Brivio:2017vri} \\ 
m_W & $W^\pm$ boson pole mass & \cite{Aaltonen:2013iut,Aaboud:2017svj} & \cite{Brivio:2017vri,Bjorn:2016zlr,Awramik:2003rn} \\ 
\Gamma_W & Total width of the $W^\pm$ boson & \cite{Patrignani:2016xqp} & \cite{Brivio:2017vri} \\ 
\text{BR}(W^\pm\to  e^\pm\nu) & Branching ratio of $W^\pm\to  e^\pm\nu$, summed over neutrino flavours & \cite{Schael:2013ita} & \cite{Brivio:2017vri} \\ 
\text{BR}(W^\pm\to \mu^\pm\nu) & Branching ratio of $W^\pm\to \mu^\pm\nu$, summed over neutrino flavours & \cite{Schael:2013ita} & \cite{Brivio:2017vri} \\ 
\text{BR}(W^\pm\to \tau^\pm\nu) & Branching ratio of $W^\pm\to \tau^\pm\nu$, summed over neutrino flavours & \cite{Schael:2013ita} & \cite{Brivio:2017vri} \\ 
\text{R}(W^+\to cX) & Ratio of partial width of $W^+\to cX$, $X=\bar d, \bar s, \bar b$ over the hadronic $W$ width & \cite{Tanabashi:2018oca} & \cite{Brivio:2017vri} \\ 
\text{R}_{\mu  e}(W^\pm\to \ell^\pm\nu) & Ratio of branching ratio of $W^\pm\to \mu^\pm\nu$ and  $W^\pm\to  e^\pm\nu$, individually summed over neutrino flavours & \cite{Aaij:2016qqz} & \cite{Brivio:2017vri} \\ 
\text{R}_{\tau  e}(W^\pm\to \ell^\pm\nu) & Ratio of branching ratio of $W^\pm\to \tau^\pm\nu$ and  $W^\pm\to  e^\pm\nu$, individually summed over neutrino flavours & \cite{Abbott:1999pk} & \cite{Brivio:2017vri} \\ 
A_ s & Asymmetry parameter in $Z^0\to s\bar s$ & \cite{Abe:2000uc} & \cite{Brivio:2017vri} \\ 
R_{uc}^0 & Average ratio of $Z^0$ partial widths to $u$ or $c$ pairs vs. all hadrons & \cite{Tanabashi:2018oca} & \cite{Brivio:2017vri,Freitas:2014hra} \\
\bottomrule
\end{tabularx}
\caption{The EWPO observables used in our fits, and the experimental measurements and theory implementations used in \flavio.}
\label{tab:smelli_obs_EWPT}
\end{table*}

\bibliographystyle{JHEP_MJKirk} 
\bibliography{bibliography}

\end{document}